\documentclass[aps,prl,superscriptaddress,showpacs,floatfix,twocolumn]{revtex4}

\usepackage{graphicx}% Include figure files
\usepackage{dcolumn}% Align table columns on decimal point
\usepackage{bm}% bold math
\usepackage{units}

\begin{document}

\title{Evolution of the nuclear modification factors                        
       with rapidity and centrality in d+Au collisions at $\sqrt{s_{NN}} = \unit[200]{GeV}$}

\newcommand{\bnl}           {$\rm^{1}$}
\newcommand{\ires}          {$\rm^{2}$}
\newcommand{\kraknuc}       {$\rm^{3}$}
\newcommand{\krakow}        {$\rm^{4}$}
\newcommand{\baltimore}     {$\rm^{5}$}
\newcommand{\newyork}       {$\rm^{6}$}
\newcommand{\nbi}           {$\rm^{7}$}
\newcommand{\texas}         {$\rm^{8}$}
\newcommand{\bergen}        {$\rm^{9}$}
\newcommand{\bucharest}     {$\rm^{10}$}
\newcommand{\kansas}        {$\rm^{11}$}
\newcommand{\oslo}          {$\rm^{12}$}

\author{
  I.~Arsene\bucharest, % \and
  I.~G.~Bearden\nbi, % \and
  D.~Beavis\bnl, % \and
  C.~Besliu\bucharest, % \and
  B.~Budick\newyork, % \and
  H.~B{\o}ggild\nbi, % \and
  C.~Chasman\bnl, % \and
  C.~H.~Christensen\nbi, % \and
  P.~Christiansen\nbi, % \and
  J.~Cibor\kraknuc, % \and
  R.~Debbe\bnl, % \and
  E.~Enger\oslo,  %\and
  J.~J.~Gaardh{\o}je\nbi, % \and
  M.~Germinario\nbi, % \and
  K.~Hagel\texas, % \and
  H.~Ito\bnl, % \and
  A.~Jipa\bucharest, % \and
  F.~Jundt\ires, % \and
  J.~I.~J{\o}rdre\bergen, % \and
  C.~E.~J{\o}rgensen\nbi, % \and
  R.~Karabowicz\krakow, % \and
  E.~J.~Kim{$\rm^{1,11}$}, % \and
  T.~Kozik\krakow, % \and
  T.~M.~Larsen\oslo, % \and
  J.~H.~Lee\bnl, % \and
  Y.~K.~Lee\baltimore, % \and
  S.~Lindal\oslo, % \and
  R.~Lystad\bergen,	
  G.~L{\o}vh{\o}iden\oslo, % \and
  Z.~Majka\krakow, % \and
  A.~Makeev\texas, % \and
  M.~Mikelsen\oslo, % \and
  M.~Murray{$\rm^{8,11}$}, %
  J.~Natowitz\texas, % \and
  B.~Neumann\kansas, % \and
  B.~S.~Nielsen\nbi, % \and
  D.~Ouerdane\nbi, % \and
  R.~P\l aneta\krakow, % \and
  F.~Rami\ires, % \and
  C.~Ristea\bucharest, % \and
  O.~Ristea\bucharest, % \and
  D.~R{\"o}hrich\bergen, % \and
  B.~H.~Samset\oslo, % \and
  D.~Sandberg\nbi, % \and
  S.~J.~Sanders\kansas, % \and
  R.~A.~Scheetz\bnl, % \and
  P.~Staszel\nbi, % \and
  T.~S.~Tveter\oslo, % \and
  F.~Videb{\ae}k\bnl, % \and
  R.~Wada\texas, % \and
  Z.~Yin\bergen, and
  I.~S.~Zgura\bucharest\\ % \and
  The BRAHMS Collaboration \\ [1ex]
  \bnl~Brookhaven National Laboratory, Upton, New York 11973 \\
  \ires~Institut de Recherches Subatomiques and Universit{\'e} Louis
  Pasteur, Strasbourg, France\\
  \kraknuc~Institute of Nuclear Physics, Krakow, Poland\\
  \krakow~Smoluchkowski Inst. of Physics, Jagiellonian University, Krakow, Poland\\
  \baltimore~Johns Hopkins University, Baltimore 21218 \\
  \newyork~New York University, New York 10003 \\
  \nbi~Niels Bohr Institute, Blegdamsvej 17, University of Copenhagen, Copenhagen 2100, Denmark\\
  \texas~Texas A$\&$M University, College Station, Texas, 17843 \\
  \bergen~University of Bergen, Department of Physics, Bergen, Norway\\
  \bucharest~University of Bucharest, Romania\\
  \kansas~University of Kansas, Lawerence, Kansas 66045 \\
  \oslo~University of Oslo, Department of Physics, Oslo, Norway\\
  %$^+ Deceased$
 }

\date{\today}% It is always \today, today,
             %  but any date may be explicitly specified

\begin{abstract}

  We report on a study of the transverse momentum dependence of
  nuclear modification factors $R_{dAu}$ for charged hadrons produced
  in deuteron + gold collisions at $\sqrt{s_{NN}}=\unit[200]{GeV}$, as
  a function of collision centrality and of the pseudorapidity ($\eta
  = 0,1,2.2,3.2 $) of the produced hadrons. We find a significant and
  systematic decrease of $R_{dAu}$ with increasing rapidity. The
  mid-rapidity 
  enhancement and the forward rapidity suppression
  are more pronounced in central collisions relative to peripheral
  collisions. These results are relevant to the study of the possible
  onset of gluon saturation at energies reached at BNL RHIC.

\end{abstract}

% PACS, the Physics and Astronomy
                                   % Classification Scheme.
%\keywords{Suggested keywords}     %Use showkeys class option if keyword
                                   %display desired
\maketitle

Studies of deep inelastic scattering of leptons on
protons and nuclei have revealed a large component of gluons 
with small--$x$ (i.e. fraction of the nucleon momentum)
that appears to diverge with
decreasing $x$~\cite{HERAdata}. However, it has also been suggested
that the density of gluons remains finite due to the increased role of
gluon-gluon correlations (`gluon fusion'), forcing an upper limit on
the total number of highly delocalized small--$x$ gluons~\cite{Levin,Mueller:wy}.
Phenomenological descriptions of DESY {\it ep} collider HERA and Fermilab
results~\cite{Golec,Lambda} based on gluon saturation appear to
successfully describe the data. Consequently, nuclei at high
energies may be thought of as highly correlated systems of small--$x$
gluons.
A QCD based theory for dense small--$x$ systems, termed the
Color Glass Condensate (CGC) 
has been developed                                                 
~\cite{McLerranVenu}.

Collisions between hadronic systems at a center-of-mass energy
$\sqrt{s_{NN}}=\unit[200]{GeV}$ at the BNL Relativistic Heavy Ion Collider
(RHIC) provide a window on the small--$x$ gluon distributions of swiftly
moving nuclei. In particular, collisions between deuterons and gold
nuclei in which hadrons with $p_{T}>1$GeV/{\it c}, mostly produced by quark--gluon interactions,
are detected close to the deuteron beam direction, 
allow for probing the small--$x$
components of the wave function of the gold nuclei.
It has been predicted 
that gluon saturation effects will manifest themselves as a
suppression in the transverse momentum distribution below a value that sets 
the scale of the effect 
~\cite{McLerranVenu,Dumitru:2001jn}. The
transverse momentum scale for the onset of gluon saturation depends on
the gluon density and thus on the number of nucleons,
and is connected with the rapidity $y$ of measured particles by $Q^2_s
\sim A^{1/3} e^{\lambda y}$, where $\lambda \sim 0.2-0.3$ is
obtained from fits to HERA data.
Thus saturation effects are most evident at large $y$ or
pseudorapidity $\eta$, i.e. at small angles relative to the beam direction. At RHIC energies
and at mid-rapidity the saturation scale for Au ions is expected to be $\sim 2\ GeV^2$
~\cite{McLerranVenu,Dumitru:2001jn}.

\begin{figure*}[!ht]
  \resizebox{0.99\textwidth}{!}
  {\includegraphics{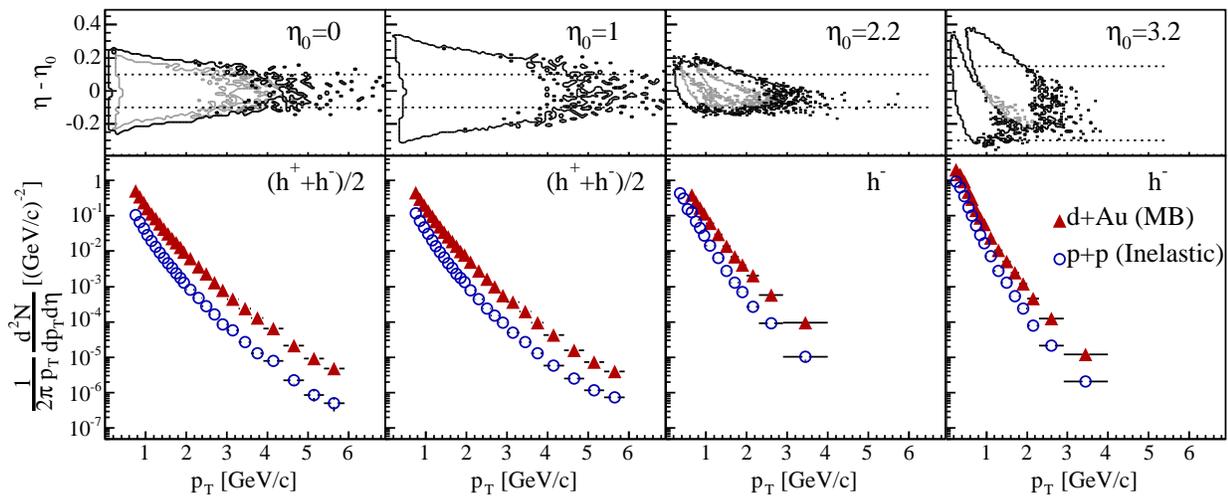}}
  \caption{\label{fig:spectra} Bottom row: Invariant yield
    distributions for charged hadrons produced in {\it d}+Au and {\it p+p}
    collisions at $\sqrt{s} = \unit[200]{GeV}$ at pseudorapidities
    $\eta=0,1.0,2.2,3.2$. Horizontal lines indicate bin width. 
    Top row: Outline of the data samples
    in $\eta$ vs $p_T$ collected with the BRAHMS spectrometers at the various
    angle and magnetic field
    settings (shifted in $\eta$), gray lines are used to indicate the overlap of the settings.
The dotted lines indicate the cuts applied
    in $\eta$.}
\end{figure*}

We report measurements of transverse momentum spectra of hadrons from
{\it p+p} and {\it d}+Au collisions at $\sqrt{s_{NN}}=\unit[200]{GeV}$ in four
pseudorapidity ranges around $\eta = 0$, 1, 2.2, and 3.2 (rapidities of 
the deuteron and gold nuclei are $+5.4$ and
$-5.4$, respectively).  
A pion with transverse momentum $p_T=\unit[2]{GeV/{\it c}}$, at these 
rapidities, probes the gluon distribution in the gold nuclei down to $x$ 
values that range from $0.01$ at $\eta = 0$  to $4\times10^{-4}$ at 
$\eta = 3.2$.  We compare the yields 
from {\it d}+Au collisions to those from {\it p+p}, scaled by the average number of
binary collisions $\langle N_{coll} \rangle$ in a {\it d}+Au event. Results around mid-rapidity
 have previously been reported~\cite{RHICdA,BRAHMSdA}.

The data presented here 
were collected
with the BRAHMS detector system~\cite{BRAHMSNIM}, consisting of event characterization
detectors and two rotatable magnetic spectrometers: the Forward Spectrometer
(FS) and the Mid-Rapidity Spectrometer (MRS).
 For the present studies the MRS was positioned at 90 and 40
degrees and the FS at $12$ and $4$ degrees with respect to the deuteron 
direction. 
The experimental method and analysis techniques
employed here are similar to those used for the study of {\it d}+Au and
Au+Au collisions~\cite{BRAHMSdA}, except that the present data at
$\eta = 2.2$ and $\eta = 3.2$ were analyzed using the front part of
the FS detector systems only.
The minimum bias trigger is estimated to select $91\% \pm 3\%$ of the
{\it d}+Au inelastic cross section and $71\% \pm 5\%$ of the total inelastic
proton-proton cross section of 41 mb. 
 The {\it p+p} yields have been corrected for trigger bias. 
Our trigger selects non-single-diffractive events and we estimate using 
the PYTHIA model that the correction should be 
$13\pm5\%$, approximately independent of $p_T$ and $\eta$.

Figure~\ref{fig:spectra} shows
the invariant yields of charged hadrons obtained from {\it d}+Au collisions
and {\it p+p} collisions in narrow pseudorapidity intervals around
$\eta=0,1,2.2$ and $3.2$. 
The spectra at $\eta=2.2$ and 3.2 are for negative
hadrons only.  
Each distribution was
constructed from independent measurements at several magnetic field
settings, as shown in the upper panels of Fig. \ref{fig:spectra} and is corrected for the 
spectrometer acceptance and
tracking efficiency. The FS acceptance ranges from 2 to 4\% 
and is known with an accuracy that ranges from 3 to 5\% of those values.  No corrections for the finite momentum
resolution, binning effects, absorption or weak decays have been applied; the 15\%  
systematic error on the spectra includes the contribution from these effects.
However, at the overlap between field settings at 4 degrees (between 1 and 2 GeV/{\it c}), 
the systematic error is 20\%.  
The momentum resolution of the spectrometers at
the maximum magnetic field setting is $\delta p/p=0.0077p$ for the MRS
and $\delta p/p=0.0018p$ for the FS,
(with {\it p} in GeV/{\it c}).

\begin{figure*}[!ht] 
\resizebox{0.99\textwidth}{!}
          {\includegraphics{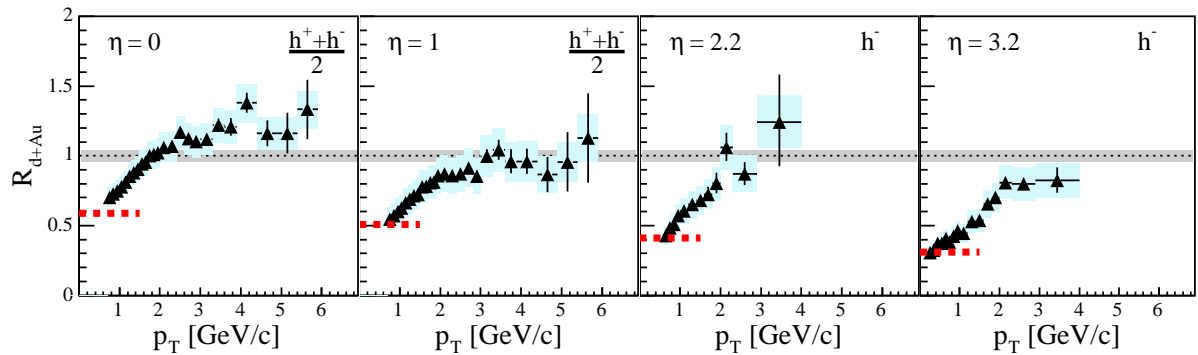}}
\caption{\label{fig:ratio}Nuclear modification factor for charged
  hadrons at pseudorapidities $\eta=0,1.0,2.2,3.2$. One standard
  deviation statistical errors are shown with error bars. Systematic
  errors are shown with shaded boxes 
  with widths set by the bin sizes.    
  The 
  shaded band around
  unity indicates the estimated error on the normalization to $\langle N_{coll} \rangle$. 
  Dashed lines at $p_T<1.5$ GeV/{\it c} show the normalized charged particle 
  density ratio $\frac{1}{\langle
  N_{coll}\rangle}\frac{dN/d\eta(Au)}{dN/d\eta(pp)}$.}
\end{figure*}

Figure~\ref{fig:ratio} compares the {\it d}+Au spectra to {\it p+p}
distributions using the nuclear modification factor defined by:
\begin{equation}
R_{dAu} \equiv \frac{1}{\langle N_{coll} \rangle}
        \frac{d^2N^{d+Au}/dp_Td\eta}{d^2N^{p+p}_{inel}/dp_Td\eta}. \label{equation1}
\end{equation} 
For the  minimum--bias sample We estimate the mean number of binary collisions $\langle N_{coll} \rangle = 7.2 \pm 0.3$, using the
HIJING v.1.383 event generator~\cite{HIJING} and a GEANT based
Monte-Carlo simulation of the experiment.  
In these  ratios most systematic errors cancel.  Remaining
systematic errors arising from variations in collision vertex
distributions, trigger efficiencies and background conditions etc. are
estimated to be less than 10\% at $\eta=0$ and less than 15\% at all 
other angle settings. 

\begin{figure*} [bht]
  \resizebox{1.1\linewidth}{!}  
  {\includegraphics{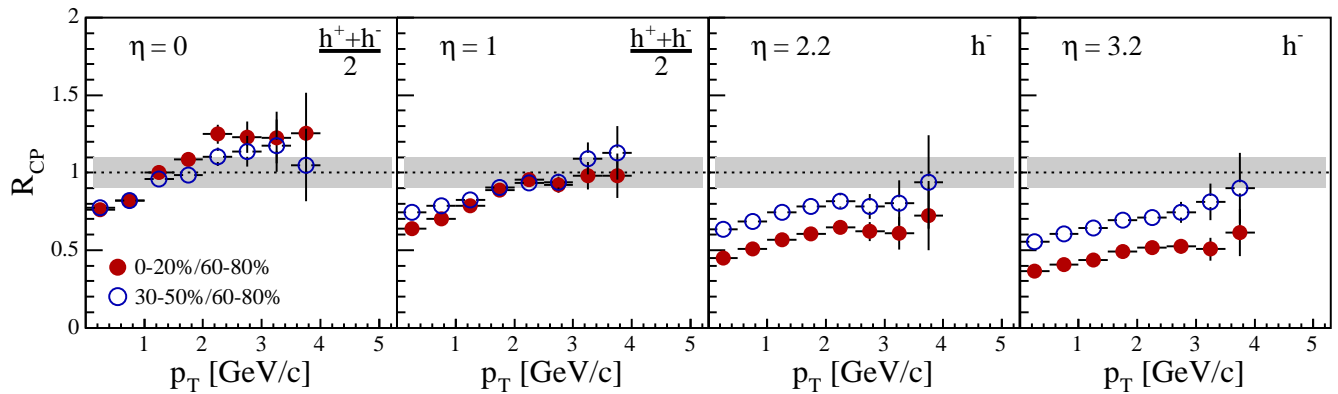}}
  \caption{\label{fig:centrality} Central (full points) and
    semi-central (open points) $R_{cp}$ ratios (see text for details)
    at pseudorapidities $\eta=0,1.0,2.2,3.2$. Systematic errors ($\sim5\%$) 
    are smaller than the symbols.}
\end{figure*}

Figure \ref{fig:ratio} reveals a clear variation of the $R_{dAu}$
as a function of pseudorapidity.  At mid-rapidity,
$R_{dAu}(p_T>\unit[2]{GeV/{\it c}}) > 1$ shows a Cronin type  enhancement ~\cite{Cronin} as
compared to the binary scaling limit. 
At $\eta=1$ the Cronin peak is not present and at
more forward rapidities ($\eta= 3.2$) the data show a suppression at all $p_T$.
The values of the $R_{dAu}$ ratios 
at low $p_T$ are observed to be similar to the ratio of charged-particle pseudorapidity 
densities in {\it d}+Au ~\cite{brahms_daumult} and {\it p+p} ~\cite{Alner:1986xu} collisions  $\frac{1}{\langle
N_{coll}\rangle}\frac{dN/d\eta(Au)}{dN/d\eta(pp)}$ shown in
Fig.~\ref{fig:ratio} with dashed lines at $p_T <1.5 GeV/{\it c}$ .

Figure \ref{fig:centrality} shows the ratio $R_{cp}$ of yields from
collisions of a given centrality class (0-20\% or 30-50\%) to yields
from more peripheral collisions (60-80\%), scaled by the mean number
of binary collisions in each sample.  The centrality selection is based on charged
particle multiplicity in the range $-2.2 < \eta < 2.2$ as described in
\cite{brahms_daumult}.  
Since the peripheral collisions are similar to {\it p+p},
the $R_{cp}$ is dominated by the nuclear effects in the more
central collisions, making  
the nuclear modification independent of the {\it p+p} reference spectrum.  
The data from the different centrality classes are obtained from the same
collider run. The ratios shown in Fig.~\ref{fig:centrality} are
therefore largely free of systematic errors associated with
run--by--run Collider and detector performance, and wide $\eta$ bins can be used for
each spectrometer setting. 
In contrast, the ratios shown in Fig. \ref{fig:ratio} must be constructed
from two collider runs with different species. Smaller $\eta$ bins must then 
be used in order to include detailed acceptance corrections leading to larger 
fluctuations.
The dominant systematic error in the $R_{cp}$ ratios 
comes from the determination of $\langle N_{coll}\rangle$ in the
centrality bins. The shaded bands in Fig. \ref{fig:centrality} indicate the uncertainty
in the calculation of $\langle N_{coll}\rangle$ in the peripheral collisions (12\%).  
We estimate the mean number of binary collisions in the
three centrality classes to be $\langle
N_{coll}^{0-20\%}\rangle=13.6\pm0.3$, $\langle
N_{coll}^{30-50\%}\rangle=7.9\pm0.4$ and $\langle
N_{coll}^{60-80\%}\rangle=3.3\pm0.4$.

\begin{figure} [!ht]
\vspace{-5mm}
\resizebox{0.40\textwidth}{!}
           {\includegraphics{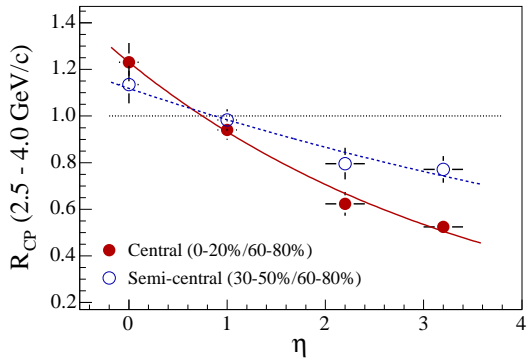}}
           \vspace{-3mm}
\caption{\label{fig:rcpsummary} Evolution of the
  central/peripheral (full points) and semicentral/peripheral (open
  points) $R_{cp}$ ratios on pseudorapidity.}
\end{figure}

There is a substantial
change in $R_{cp}$ between $\eta = 0$ and the forward rapidities.  At
low pseudorapidity, the central--to--peripheral collisions ratio is
larger than the semicentral--to--peripheral ratio, suggesting the
increased role of Cronin like multiple scattering effects in the more
violent collisions. Conversely, at forward pseudorapidities the more
central ratio is smallest
indicating a suppression mechanism
that depends on the centrality of the collision.
In Fig.~\ref{fig:rcpsummary} 
we show  $R_{cp}$ for the transverse momentum interval
$p_T=\unit[2.5-4.0]{GeV/{\it c}}$. A fit to 
$R_{CP} \sim e^{\alpha \eta}$
yields 
$\alpha = -0.28 \pm  0.03$ for the central--to--peripheral ratio, a similar 
$ \eta$ dependence as $Q^2_s$ from HERA \cite{Lambda}, 
 $\alpha = -0.13 \pm  0.03$ for the semicentral--to--peripheral ratio.

The observed suppression of 
yield in {\it d}+Au collisions
(as compared to {\it p+p} collisions) has been qualitatively predicted by
several authors~\cite{KKT,Wiedemann,Jalilian-Marian:2003mf} within the
framework of gluon saturation that includes the effects of  
`quantum evolution' with rapidity. However, no detailed
numerical predictions are yet available. These approaches also predict
the observed centrality dependence of the suppression at different  pseudorapidities.
Other authors~\cite{Vitev:2003xu,XWang} have based their
predictions 
on the two component microscopic HIJING model
that includes a parametrization of perturbative QCD and string breaking
as a mechanism to account for soft coherent particle production, and `gluon shadowing'
as a method for reducing the number of
effective gluon--gluon collisions.
 The HIJING model has been shown
to give a good description of the overall charged particle
distribution in {\it d}+Au collisions~\cite{brahms_daumult,phobos_mult}, and
thus the low--$p_{T}$ behavior of $R_{dAu}$ with pseudorapidity.

In summary, we observed a significant reduction of the yield of
charged hadrons 
measured in {\it d}+Au collisions, as compared to scaled {\it p+p} collisions at
forward pseudorapidities. This suppression for $p_T>\unit[2]{GeV/{\it c}}$, is absent at
mid-rapidity ~\cite{RHIC_auau_supp,BRAHMSdA}, increases smoothly 
as the
difference in rapidity between the detected particles and the  
gold ion increases.                                            
Also, the
change from mid-- to forward rapidities is stronger for central
collisions than for semicentral collisions, indicating 
a dependence on the geometry of the collision. Such effects are consistent with the onset of saturation in the Au nuclei gluon density
at small--$x$ values which modifies the shapes and magnitudes of $R_{dAu}$ and $R_{cp}$ at all
transverse momenta.

 These results highlight  
opportunities for studying saturation phenomena in nuclei at RHIC.

%\begin{acknowledgments}
This work was supported by 
the Office of Nuclear Physics of the U.S. Department of Energy, 
the Danish Natural Science Research Council, 
the Research Council of Norway, 
the Polish State Committee for Scientific Research (KBN) 
and the Romanian Ministry of Research.

%\end{acknowledgments}

\bibliographystyle{apsrev}
\bibliography{apsrev}% Produces the bibliography via BibTeX. RD was samp changed to rev

\end{document}